\pdfoutput=1
\documentclass{emulateapj}
\usepackage{apjfonts}
\usepackage{amsmath,graphicx}
\newcommand{\beq}{\begin{equation}}
\newcommand{\eeq}{\end{equation}}
\newcommand{\bea}{\begin{eqnarray}}
\newcommand{\eea}{\end{eqnarray}}

\begin{document}

\title{Accretion onto Intermediate-Mass Black Holes in Dense Protogalactic Clouds}
\author{
Milo\v s Milosavljevi\'c,
Sean M. Couch, and
Volker Bromm
}
\affil{Department of Astronomy, University of Texas, 1 University Station C1400, Austin, TX 78712.}
%\altaffiltext{1}{Department of Astronomy, University of Texas, 1 University Station C1400, Austin, TX 78712.}
\righthead{IMBH ACCRETION IN DENSE CLOUDS}
\lefthead{MILOSAVLJEVI\'C, COUCH, \& BROMM}

\begin{abstract}

We present the first results from two-dimensional simulations of radiatively-efficient accretion of metal-free gas 
onto intermediate-mass black holes.  We fix the shape of the spectral energy distribution of the radiation produced near the event horizon and study the structure of the irradiated 
low-angular-momentum accretion flow over three orders of magnitude in
radius from the black hole,
$10^{14}-10^{17}\textrm{ cm}$ for a $100~M_\odot$ black hole.  The
luminosity of the central source is made to be proportional
to  the rate at which gas accretes across the inner boundary, which we
set just inside the sonic radius.  
We find that photoionization heating and
radiation pressure modify the structure of the flow.  When the ambient gas density is 
$10^7\textrm{ cm}^{-3}$,
accretion is intermittent and on average reduced to $32\%$ of the Eddington-limited rate, two orders of magnitude below the ``Bondi'' rate
evaluated ignoring radiation, in agreement with simplified theoretical models.  Even if the vicinity of the black hole is supplied with high density gas, accretion is rendered inefficient through heating and radiation pressure.  


\end{abstract}

\section{Introduction}
\label{sec:introduction}

Known stellar-mass and intermediate-mass black hole candidates accrete through mass transfer from a
stellar companion without exception. 
Rapidly-accreting massive black hole candidates, such as in Seyfert
galaxies and quasars, appear to be situated
close to dusty, molecular, and highly inhomogeneous gas
reservoirs, where cold, dense clumps are in free fall or
are embedded in a hot, ionized, and tenuous
confining medium.  In neither case does the accretion flow
resemble the laminar flow of the Bondi-Hoyle-Littleton solution.
There is a distinct possibility, however,
that massive black hole progenitor ``seeds,'' with masses $M_{\rm
  bh}\gtrsim 100~M_\odot$ \citep{Madau:01}, did pass through an early
phase of quasiradial accretion in protogalaxies, prior to significant
dust enrichment and the emergence of a pervasive cold phase in the
interstellar medium \citep[e.g.,][]{Volonteri:05,Alvarez:06,Alvarez:09,Johnson:07,Pelupessy:07,DiMatteo:08,Greif:08}. This was the period immediately following the
onset of atomic cooling, the epoch when protogalaxies experienced rapid growth
due to cosmic infall and yet had potential wells too shallow to
retain all of their newly synthesized metals.  

Since the accretion rate in quasiradial, Bondi-like accretion is
proportional to gas density, Bondi-like accretion presents an appealing
formation process for massive black holes: given the presence of
sufficiently dense gas in a protogalactic nucleus,
Eddington-limited accretion and mass exponentiation on a Salpeter
time scale would proceed; stellar-size seed black holes would
reach masses $\gtrsim 10^9 M_\odot$ 
inferred in high-redshift quasars in the cosmic time
available to them \citep[e.g.,][]{Haiman:01,Volonteri:03,Tanaka:08}.  A potential flaw in this reasoning was noticed early,
namely, that in the event of radiatively-efficient accretion,
radiative heating and radiation pressure may reduce
the accretion rate \citep{Shvartsman:71,Buff:74,Hatchett:76,Ostriker:76,Cowie:78,Stellingwerf:82,Begelman:85,Ricotti:08,Alvarez:09}.  The possibility of a transition to
radiative-feedback-driven temporal intermittency, 
which has been evident in one-dimensional simulations of
Compton-heated accretion onto massive black holes in 
galaxy clusters \citep{Ciotti:01,Ciotti:07,Sazonov:05},
has further complicated the assessment of quasiradial accretion's
viability as a massive black hole formation process.  If the accretion
is indeed intermittent and episodic, 
how long is the duty cycle, how large is the
average accretion rate, and does it permit evolution
of seed black holes into quasars? Does episodic accretion proceed
in bursts, where episodes of Eddington-limited rapid accretion
 alternate with near-complete radiative quenching and outflow, as
one-dimensional models appear to suggest, or does the multidimensional
nature of realistic irradiated accretion flows imply a damping of the
fluctuations and convergence to a quasi-steady state characterized by a reduced accretion rate?

Here, we present the first results from our program to
investigate radiatively-efficient 
accretion onto black holes embedded in a realistic galactic gaseous
medium with the help of high-resolution hydrodynamic simulations. We
do not attempt to simulate fluid flow and radiation emission
mechanisms near the event horizon of the black
hole; instead, we simply fix the shape of the
spectral energy distribution of the radiation produced
near the event horizon and simulate the irradiated fluid flow on radial 
scales spanning the sonic radius, where the photoheated fluid falls supersonically toward the black hole, and a much larger
radius in the gas cloud that is entirely unaffected by the black
hole's radiative output, where boundary conditions for the accretion
flow are set.  This work is organized as follows. In
\S~\ref{sec:numerical} we concisely describe our numerical algorithm.  
A more detailed description and a suite of tests will be
provided in a longer follow-up paper (Couch et al., in preparation).  
In \S~\ref{sec:results} we
describe our results with particular focus on the
intermittency of the accretion flow.  In \S~\ref{sec:conclusions} we
compare our results to analytical models \citep[][and references
therein]{Milosavljevic:08} and discuss the outlook for massive black hole
formation through quasiradial accretion.

\section{Numerical Algorithm}
\label{sec:numerical}

Here we describe numerical methodologies that we have implemented in
the parallel adaptive-mesh-refinement (AMR) code FLASH \citep{Fryxell:00},
version 2.5, to facilitate our 
simulations of low-angular momentum accretion 
flow with a central ionizing source. The hydrodynamic solver
used was the piecewise parabolic method (PPM), which is a higher-order
Godunov method. Contact steepening that sharpens contact
discontinuities in the PPM module was disabled after we noticed that
it led to spurious flow near stationary ionization fronts.  By default, FLASH does not conserve
angular momentum in 2D cylindrical simulations. To simulate fluid
motion in the azimuthal direction, we
introduce the specific angular momentum $l\equiv rv_\phi$ as an
additional conserved ``mass
scalar'' quantity that is advected with the fluid.  Equivalence with
3D cylindrically symmetric equations of fluid dynamics is achieved by
adding the centrifugal force to the total acceleration, ${\bf a}={\bf
  a}_{\rm grav} + {\bf a}_{\rm rad} +  l^2\hat {\bf R}/R^3$.

\subsection{The Central Hole and the Central Ionizing Source}
\label{sec:hole}

%Here we describe our excision of the central hole and the implementation of a central ionizing source. 
The simulations were carried out in cylindrical symmetry on a
rectangular $(R,z)$ grid in two spatial dimensions, where $R$ is the distance from the axis of
symmetry.  In what follows, $r$ will denote the
distance from the origin, $r^2=R^2+z^2$.  A small spherical region
$r<r_{\rm hole}$ was excluded from the hydrodynamical update; we refer
to this region as the ``hole.''  A unidirectional
``outflow'' boundary condition, allowing only outflow from the simulated spatial domain and inflow into the central hole, was imposed at 
$r=r_{\rm hole}$. The gravitational acceleration was set to that of a Newtonian point mass located at the origin of the grid ${\bf a}_{\rm grav}=-G M_{\rm hole}\hat {\bf r}/ r^2$.  
The mass of the material flowing into the hole was added
to the point mass $M_{\rm hole}$.  Self gravity of the fluid was ignored because, typically, the mass of the photoionized gas around the black hole was less than $1\%$ of the mass of the black hole.

An isotropic source of ionizing radiation was placed at the origin of
the grid. The total luminosity of the ionizing source between
energies $h\nu_{\rm min}=13.6\textrm{ eV}$ and $h\nu_{\rm
  max}=100\textrm{ keV}$ was set to be proportional to the accretion
rate into the hole, 
\begin{equation}
\int_{\nu_{\rm min}}^{\nu_{\rm max}} L_\nu d\nu = \epsilon \dot M_{\rm
  hole} c^2 ,
\end{equation}
where $\epsilon$ is a radiative efficiency which was uniformly
set to $0.1$.  The radiative spectral energy distribution was set
to the power law $L_\nu \propto \nu^{-3/2}$ and was calculated in 40
logarithmically-spaced energy bins.

\subsection{Composition, Chemistry, and Thermodynamics}
\label{sec:chemistry}

%Here we discuss composition of the accretion flow, and the photochemical and thermodynamic processes in the flow. 
We simulated a six-species primordial fluid containing hydrogen
and helium and their ions: H,
H$^+$, He, He$^+$, He$^{++}$, and $e^{-}$.   Abundances
of the six species were integrated in time with a chemical network
that includes collisional ionization of H, He, and He$^+$, ``case B''
recombination of H$^+$, He$^+$, and He$^{++}$, bidirectional charge exchange between H
(H$^+$) and He$^+$ (He), as well as primary and secondary
photoionization of H and He.  Secondary photoionization and heating by
fast photoelectrons were implemented in accordance with the results of
\citet{Shull:85} and \citet{Dalgarno:99}. Photoionization and photoheating rates were calculated in
a photon-conserving manner (see \S~\ref{sec:radiation}) from the local,
extincted radiative fluxes.

Photoionization heating and collisional cooling of the fluid were
operator-split from the hydrodynamic update. Cooling processes that
were taken into account include Bremsstrahlung, collisional ionization
cooling of H, He, and He$^+$, recombination cooling of H$^+$, He$^+$,
and He$^{++}$, dielectronic recombination cooling of He$^+$,
collisional excitation cooling of H and He$^+$, and Compton cooling to
the cosmic microwave background.  At the high gas densities in the
simulations, the Compton cooling was negligible.

Number densities of all six species and the temperature of the gas
were advanced 
simultaneously during the chemical and thermodynamic update 
with the Bulirsch-Stoer-type semi-implicit extrapolation mid-point method 
of \citet{Bader:83}. Our implementation of the method closely
follows that provided in the GNU Scientific Library \citep{Galassi:06}. 
This method utilizes the Jacobian matrix
of the partial derivatives of the right-hand side of our seven coupled
ordinary differential equations. We evaluate the partial derivatives
analytically.  Special care was taken to accurately integrate the
abundance of the neutral and partially ionized fraction in a
strongly photoionized gas. We carried the Richardson extrapolation
sequence to four points, 
which was sufficient for numerical stability.

In the operator-split alternating hydrodynamical and
chemothermodynamic update, the hydrodynamic step was globally
constrained to not exceed $10\%$ of the minimum electron abundance
change time scale, $\Delta t_{\rm hydro}\leq 0.1\ n_e /|\dot n_e|$, in
combination with the usual Courant limit. The chemothermodynamic update is
carried out in single or in multiple and possibly much shorter 
Bader-Deuflhard steps. The latter are individually
limited in every cell by the time 
scale on which any of the six abundances and the temperature evolves,
$\Delta t_{\rm BD}\leq \min \{ 0.1\ n_i/|\dot n_i| , 0.1\ T/\dot T,
\Delta t_{\rm hydro}\}$, where $i={\rm H},...,e^-$ \citep[cf. e.g.,][]{Whalen:06}.

\subsection{Line-of-Sight Radiative Transfer and Radiation Pressure}
\label{sec:radiation}

Here, we discuss our implementation of 
multifrequency photon-conserving line-of-sight radiative transfer and the resulting
radiation pressure force. We calculate the line-of-sight optical
depth $\tau_\nu^{(k)}$ from the central source to the faces of any
rectangular zone along $N_{\rm ray}=4$ separate rays, $k=1,...,N_{\rm ray}$; the rays are equally
spaced in angle as seen from the central source. We also calculate the
optical depth to absorption within the zone along each of the rays
$\hat \tau_\nu^{(k)}$. The effective flux at the zone center is
calculated via \citep[see, e.g.,][and references therein]{Mellema:06}
\beq
\label{eq:photon_conserving}
F_\nu = \frac{L_\nu}{4\pi r^2} 
\frac{\sum_{k=1}^{N_{\rm rays}} \exp (-\tau_\nu^{(k)}) [1-\exp (-\hat
    \tau_\nu^{(k)})]}{\sum_{k=1}^{N_{\rm rays}} \hat \tau_\nu^{(k)}} .
\eeq
The total photon absorption rates per chemical species $i$
within the zone are then calculated via
$\Gamma_{i,\nu} = \sigma_{i,\nu} n_i V F_\nu/h\nu$, where $V$ is
the zone volume.  
In the limit in which the zone is very optically thick, $\hat
\tau_\nu^{(k)}\gg 1$, we obtain
\beq
\label{eq:optically_thick}
\sum_i \Gamma_{i,\nu} \approx (h\nu)^{-1} F_\nu \langle
  \exp(-\tau_\nu^{(k)})\rangle \Omega ,
\eeq
where $\Omega$ is the angular size of the zone as seen from the central
source, and the extinction external to the zone is averaged over
rays. The approximation in equation (\ref{eq:optically_thick}) becomes exact in the
limit $N_{\rm rays}\rightarrow \infty$, as it has to be to conserve 
photons. 

We calculate radiation pressure associated with photoionization and
Thomson scattering. The radiation pressure force is added to the
gravitational force of the central point mass.  We ignore line resonance radiation altogether and also ignore any diffusion of ionizing radiation \citep[see, e.g.,][]{Ritzerveld:05}.

\subsection{Mesh Refinement and Initial Conditions}
\label{sec:amr_initial}

We simulate a cylindrical domain in the upper hemisphere $0 < (R,z) <
1\textrm{ pc}$ with 18 levels of mesh refinement. Since each AMR block
contains $8\times8$ zones, the maximum spatial resolution in the
simulation is $\Delta x\sim 3\times10^{12}\textrm{ cm}$.  We require that the
central hole, with radius $r_{\rm hole}=10^{14}\textrm{ cm}$, is 
always resolved at the highest level of mesh refinement. We achieve
pseudo-logarithmic gridding by capping the resolution at radius $r$ with
$\Delta x > \frac{1}{8} \eta r$ where we choose $\eta=0.1$; this
prevents use of excessive resolution far from the central hole.
The simulation domain initially contained uniform-density partially
ionized gas, with initial electron abundance of $\chi_e=0.5$, at temperature $10^4\textrm{ K}$ and density $n_{\rm H}=10^7\textrm{ cm}^{-3}$. The initial value for the central point
mass was $100~M_\odot$.

\begin{figure}
\begin{center}
\includegraphics[width=3.5in]{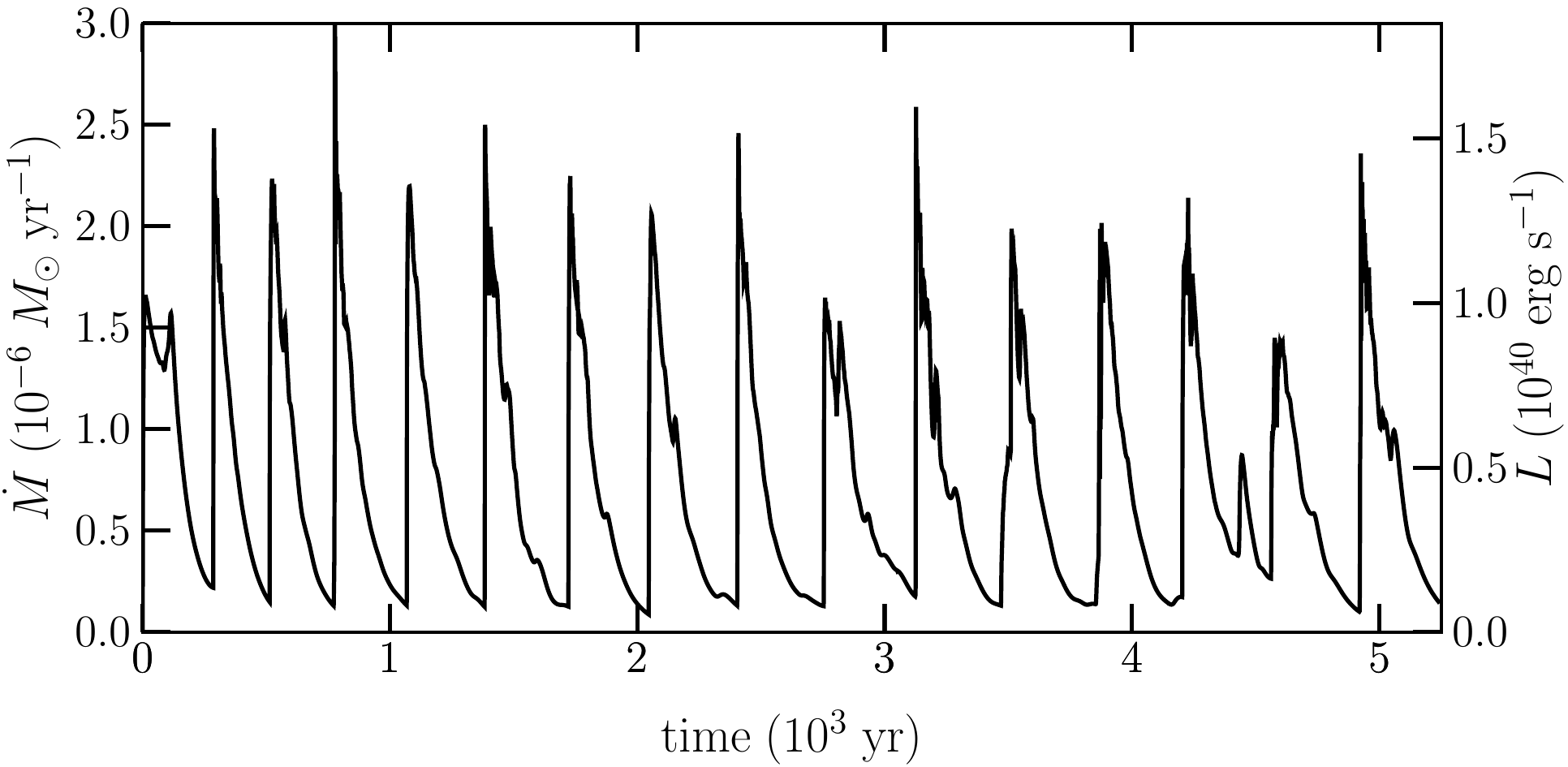}
\end{center}
\caption{Central accretion rate and radiative luminosity as a function
of time for radiative efficiency $\epsilon=0.1$.}
\label{fig:accretion_rate_luminosity}
\end{figure}

\section{Results}
\label{sec:results}

The central accretion rate, shown in Figure
\ref{fig:accretion_rate_luminosity}, oscillates between values close
to, and occasionally mildly exceeding the Eddington
limit, $\dot M_{\rm max}\sim \dot M_{\rm Edd}\sim
2\times10^{-6}M_\odot\textrm{ yr}^{-1}$, and a rate lower by an order of
magnitude, $\dot M_{\rm min}\sim 2\times10^{-7} M_\odot\textrm{ yr}^{-1}$. The
accretion is approximately periodic with  a mild trend toward an
increasing period separating consecutive peak episodes; the period
varies between $250\textrm{ yr}$ and $350\textrm{ yr}$.  Accretion
rate falloff following a maximum is roughly exponential, as is 
expected since photoionization radiation pressure in the
ionized region surrounding the black hole evacuates the accreting gas
and drives down the accretion rate \citep[see \S~3
in][]{Milosavljevic:08}.  The average accretion rate and luminosity are $\langle \dot M\rangle = 4.6\times10^{19} \textrm{ g cm}^{-3}$ and $\langle L\rangle=4.2\times10^{39}\textrm{ erg s}^{-1}$, which says that on average, the black hole accretes at $32\%$ of the Eddington limit.  The average accretion rate is still only $\sim 1\%$ of the adiabatic 
``Bondi'' accretion rate $\dot M_{\rm Bondi} =\pi (GM_{\rm bh})^2 n m_p/c_{\rm s}^{3}(\infty)$, calculated ignoring radiative feedback, for an ambient sound speed of $c_{\rm s}(\infty)=14\textrm{ km s}^{-1}$.  

\begin{figure}
\begin{center}
\includegraphics[width=3.5in]{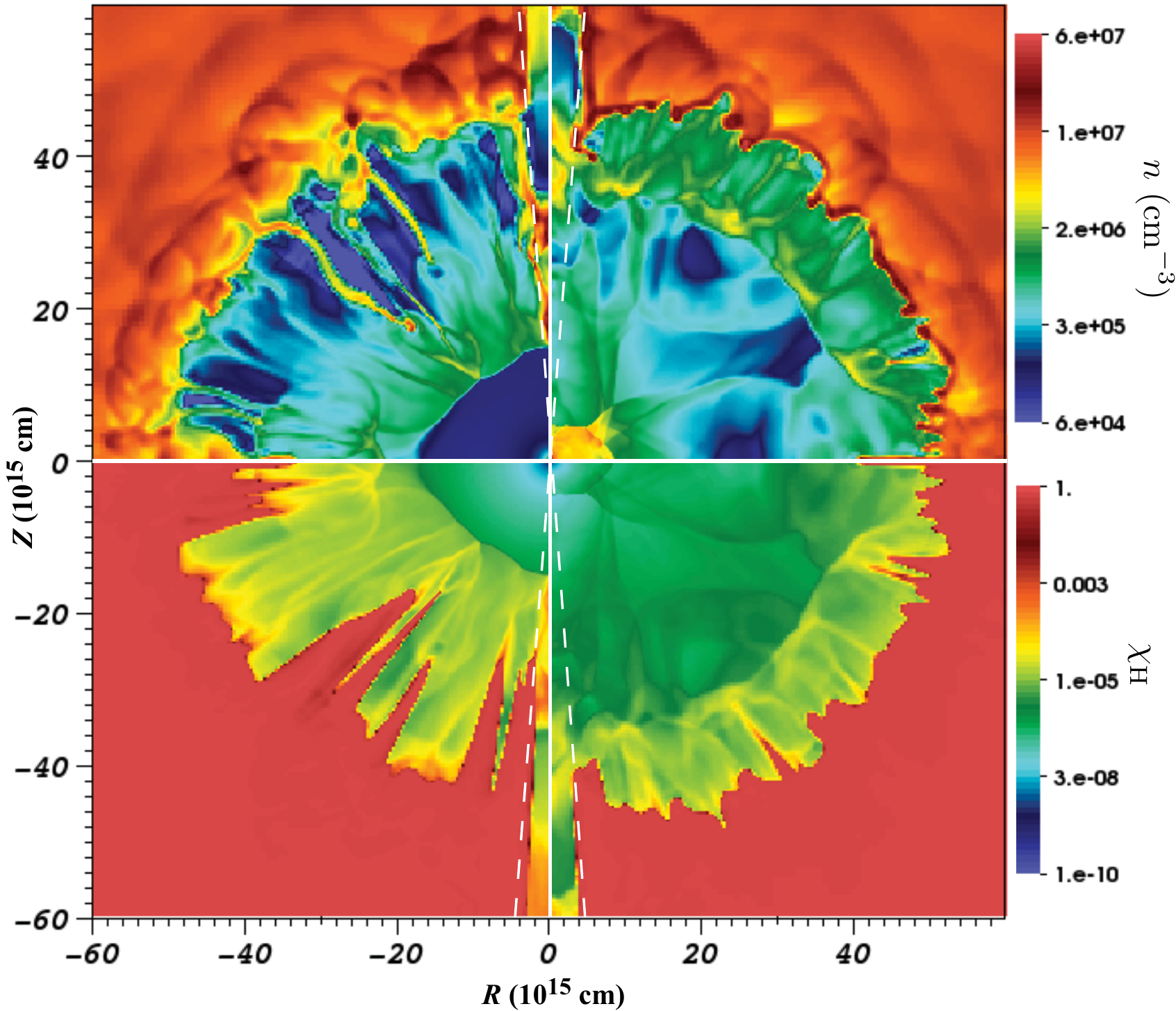}
\end{center}
\caption{Gas number density $n$ (\emph{upper panels}) and neutral fraction
  $\chi_{\rm H}$  
(\emph{lower panels}) at two separate instances around 
$t\sim 4,000\textrm{ yr}$.  The
  left panels show the flow during a central accretion
  minimum; infalling gas is clearly visible at $r\sim
  1.5\times10^{16}\textrm{ cm}$.  The right panels show the flow
  during a central accretion maximum. The structure near the central axis, in the conical region marked by dashed lines, is an artifact of our having set the optical depth to zero along the axis, for $R/|z|<\tan5^\circ$, to avoid numerical issues arising from the coordinate singularity. }
\label{fig:color_figure}
\end{figure}

During a central accretion maximum, photoionization radiation pressure drives
an outflow in the ionized gas within the \ion{H}{2} region that has
neutral fractions $\chi_{\rm H}\sim 10^{-4}-10^{-5}$
(Fig.~\ref{fig:color_figure}, \emph{lower panels}).  This leads to
rarefaction and exponential drop in central accretion.  Meanwhile, as
radiation pressure subsides, gas near the edge of the \ion{H}{2}
region accelerates inward.  This acceleration is driven by a gas pressure imbalance near the edge; the imbalance was inherited from the preceding accretion maximum when an outward radiation pressure force balanced an inward gas pressure gradient force.
  The outflow intersects with the
inflow, and the inflowing gas ultimately arrives at the edge of the
central hole and gives rise to a new accretion maximum. The longest
timescale in the cycle is the inward acceleration time, which is here a
factor of $\sim 3$ times shorter than the radial sound crossing time of the
\ion{H}{2} region with typical temperature $T_{\rm HII}\sim
4\times10^4\textrm{ K}$.

\begin{figure}
\begin{center}
\includegraphics[width=3.5in]{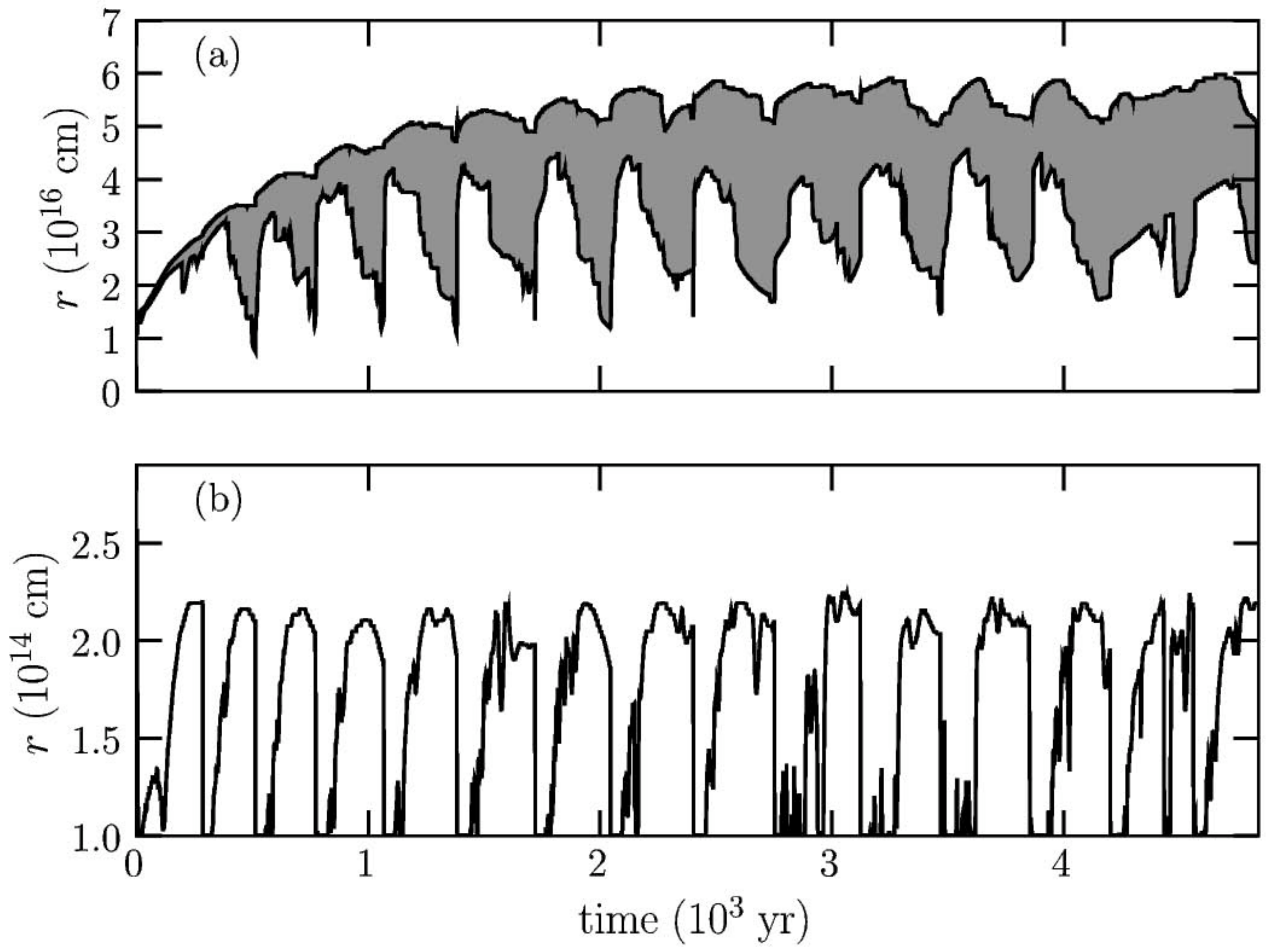}
\end{center}
\caption{Minimum radius at which neutral gas, with ionization fraction
$\chi_{\rm H^+}<\frac{1}{2}$ is found (lower boundary of the shaded
region in panel [\emph{a}]) and maximum radius at which ionized gas, with $\chi_{\rm
  H^+}>\frac{1}{2}$, is found (upper boundary of the shaded region).
Panel (\emph{b}) shows the minimum radius at which the radial inflow
is subsonic.  Because of transient shock heating in the flow, the
sonic radius, which normally resides at $r_{\rm s}\sim
2.2\times10^{14}\textrm{ cm}$, shrinks to the edge of the hole at
$r_{\rm hole}=10^{14} \textrm{ cm}$.}
\label{fig:radii}
\end{figure}

Figure \ref{fig:radii}\emph{a} shows that the radius of the \ion{H}{2} region surrounding the black hole reaches an
asymptotic value of $r_{\rm ion}\sim 6\times 10^{16}\textrm{ cm}$. This only somewhat smaller, by $40\%$, than the crude estimate obtained for hydrogen-only cloud in equation (4) of \citet{Milosavljevic:08}. The region
$r< r_{\rm ion}$ is not entirely free of neutral gas; during accretion
minima, as seen in the two left panels of Figure
\ref{fig:color_figure}, dense clumps form in the gas that has been
expelled by radiation pressure and is now falling back toward the
black hole, and the regions in the dense clumps' shadows recombine.
The range of radii that is shaded in Figure \ref{fig:radii}\emph{a}
contains a multiphase medium.  The densest neutral clumps that form near
the edges of the \ion{H}{2} region are six times
denser than the ambient gas, but this density is not sufficient to resist
photoevaporation during accretion maxima. Figure \ref{fig:radii}\emph{b} shows that the flow into the central
hole is typically supersonic with sonic radius located at $r_{\rm
  s}\sim 2\times10^{14}\textrm{ cm}$, but during accretion maxima, shock
heating in the infalling gas drives subsonic flow  to the brink of
the excised hole at $r_{\rm hole}=10^{14}\textrm{ cm}$.

\section{Conclusions and Discussion}
\label{sec:conclusions}

We have simulated accretion from a uniform, high-density ($10^7\textrm{ cm}^{-3}$)\footnote{Atomic densities have been found to reach comparable high levels in simulations of protogalactic collapse \citep[e.g.,][]{Bromm:03,Wise:07,Wise:08}.} metal-free protogalactic cloud onto an intermediate-mass black hole. At radii from the black hole smaller than the size of the excised region, $10^{14}\textrm{ cm}$, the accretion was assumed to be radiatively efficient. We find that a compact \ion{H}{2} region forms around the black hole.  The accretion is intermittent due to alternating photoionization radiation pressure-driven expulsion and external pressure-driven fallback. The average accretion rate is $\lesssim 1\%$ of the Bondi accretion rate calculated ignoring the radiation's influence, and $\sim \frac{1}{3}$ of the Eddington-limited rate. 

The numerical results are consistent with the predictions of our toy model for episodic quasiradial accretion \citep{Milosavljevic:08}, where we suggested that photoionization heating and photoionization radiation pressure within the compact  \ion{H}{2} region are the dominant accretion-suppression mechanisms.  We also suggested that Ly$\alpha$ line resonance radiation pressure, which we do not account for in the simulations presented here, should further reduce the accretion rate. The intermittency is qualitatively similar to that seen in one-dimensional simulations of the accretion of hot irradiated intergalactic medium onto black holes in central cluster galaxies by  \citet{Sazonov:05} and \citet{Ciotti:07}, with the notable caveat that photoionization radiation pressure was unimportant in that context.

The simulation in which the accreting fluid was endowed with small angular momentum corresponding to maximum circularization radius equal to $r_{\rm hole}$  exhibited similar intermittency and only a slightly reduced average accretion rate, consistent with \citet[and references therein]{Krumholz:05}.  In contrast with the nonrotating simulations that did not contain axial outflows, narrow polar outflows form that resemble some of those found by \citet{Proga:07} and \citet{Proga:08} in simulations of accretion flows irradiated by a quasar.  The accretion intermittency and inflow-outflow behavior that we observe is qualitatively similar to that seen in Proga et al., but 
additional tests are needed to ascertain whether out axial outflows are physical.

The overall goal is to combine the highly-resolved simulations presented here with large-scale cosmological simulations of how the first galaxies formed at the end of the cosmic dark ages \citep[e.g.,][]{Greif:08}.  Those simulations can resolve the central gas flows to within $\sim 0.01\textrm{ pc}$, where we can implement sub-grid prescriptions based on the calculations done here.  An important challenge will be to derive prescriptions for the feedback from black hole accretion that are applicable to a wide range of protogalactic conditions.  We will report on these efforts elsewhere (Couch et al., in preparation).

\acknowledgements

We acknowledge valuable discussions with Daniel Proga.
  The software used in this work was in part developed by the DOE-supported
ASC/Alliance Center for Astrophysical Thermonuclear Flashes at the
University of Chicago.  The authors acknowledge the Texas Advanced
Advanced Computing Center (TACC) at the University of Texas at Austin
for providing high-performance computing resources that have
contributed to this research. V.~B. and M.~M. 
acknowledge support from NSF grant AST-0708795.

\end{document}